\pdfoutput=1
\documentclass[prd,aps,nofootinbib,twocolumn,superscriptaddress,preprintnumbers,balancelastpage,longbibliography]{revtex4-1}

\usepackage{amsmath}
\usepackage{amssymb}
\usepackage{graphicx}
\usepackage{hyperref}
\usepackage{color}
\usepackage[dvipsnames]{xcolor}
\usepackage[T1]{fontenc} 
\usepackage{slashed}

\usepackage{listings}
\usepackage{color,xcolor}

\usepackage[normalem]{ulem}

\hypersetup{
     colorlinks   = true,
     citecolor    = orange,
     urlcolor     = orange,
     linkcolor    = orange
}

\newcommand{\CP}{{\sf CP}}

\allowdisplaybreaks

\newcommand{\AddrOXF}{%
Rudolf Peierls Centre for Theoretical Physics, University of Oxford, Parks Road, Oxford OX1 3PU, UK
}

\begin{document}

\title{
Searching for a dark matter induced galactic axion gradient 
}

 \author{Edward Hardy}
 \email{edward.hardy@physics.ox.ac.uk}
 \affiliation{\AddrOXF}

\author{Mario Reig}
\email{mario.reiglopez@physics.ox.ac.uk}
\affiliation{\AddrOXF}
\affiliation{Department of Physics, Royal Holloway University of London, Egham, Surrey, TW20 0EX, UK.}

\author{Juri Smirnov}
\thanks{ \href{mailto:juri.smirnov@liverpool.ac.uk}{juri.smirnov@liverpool.ac.uk}; \href{http://orcid.org/0000-0002-3082-0929}{0000-0002-3082-0929}}
\affiliation{Department of Mathematical Sciences, University of Liverpool,
Liverpool, L69 7ZL, United Kingdom}

\begin{abstract}

An ultra-light axion with {\sf CP}~violating interactions with a dark sector and {\sf CP}~preserving interactions with the visible sector can act as a novel portal between dark matter and the Standard Model. In such theories, dark matter sources an axion field extending over the entire galaxy, the gradient of which can be searched for with precise spin precession experiments. A reinterpretation of existing co-magnetometer data already constrains theories that are consistent with astrophysical bounds, and near-future experiments will begin probing well-motivated models. The required interactions can arise from a confining hidden sector without necessitating fine-tuning of the axion's mass.

\end{abstract}

\maketitle
\preprint{LTH-1373}

\section*{Introduction}

Despite compelling evidence for its existence, the nature of dark matter (DM) and its interactions with the Standard Model (SM) remain unknown. Investigation of the possible portal interactions between DM and the SM is therefore worthwhile. 
In this letter we propose a new such portal, in which the DM and the SM are connected via an ultra-light axion that has \CP~violating couplings to the DM. 
We argue that this possibility is theoretically well-motivated and can lead to observable signals in table-top experiments.

Axions (by which we mean any light pseudo-scalars) are generic in UV completions of the SM involving gauge fields in extra space-time dimensions, with string theory as the foremost example \cite{Svrcek:2006yi,Arvanitaki:2009fg,Conlon:2006tq}, see \cite{Reece:2024wrn} for a recent review. In field-theoretic realisations, an axion can couple to both the SM and a dark sector as a result of heavy matter charged under gauge groups in both sectors, while in string theory the couplings of an axion to fermions depend on the compactification \cite{Cicoli:2012sz}. 

It is plausible that the interactions of an axion with DM might be \CP~violating; the axion itself is not the DM in the scenario we consider. As a guiding analogy, recall that the QCD axion has minuscule \CP~violating interactions with SM particles only because the minimum of its potential is extremely close to the point where the strong~\CP~angle $\theta_{\rm QCD}=0$. That $\theta_{\rm QCD}\simeq 0$ is both a quirk of the SM (due to the weak interactions inducing a non-zero $\theta_{\rm QCD}$ at high loop order suppressed by powers of $G_{\rm F} \Lambda_{\rm QCD}^2\ll 1$ \cite{Georgi:1986kr}) and a theoretical problem in that contributions to the axion potential from the UV must be tiny (the quality problem \cite{Georgi:1981pu,Kamionkowski:1992mf}). One can imagine hidden sectors for which similar conditions are not fulfilled and sizable \CP~violating  interactions arise.

Meanwhile, we require that the axion does not have an anomalous coupling to QCD. We further assume that the SM is decoupled from the dominant sources of \CP~violation~\cite{DiLuzio:2020oah}, in which case is a reasonable expectation that an axion only has \CP~conserving interactions with the SM.

We therefore consider an effective Lagrangian
\begin{equation}\label{eq:lag}
    \mathcal{L}_{\text{eff}}\supset g^\chi_s \phi \Bar{\chi}\chi+c_\psi\frac{\partial_\mu \phi}{f_\phi}\Bar{\psi}\gamma^\mu\gamma^5\psi - \frac{1}{2}m_\phi^2 \phi^2\,,
\end{equation}
where $\phi$ is the axion, $f_\phi$ is its decay constant (defined such that the period of the axion field is $2\pi f_\phi$), and $m_\phi$ is the axion mass. The field $\psi$ is a generic SM fermion, and $c_\psi$ is expected not to be much bigger than $\mathcal{O}(1)$ in minimal theories. 
Meanwhile, $\chi$ is the DM, which we assume to be a fermion. In typical UV completions involving a confining hidden sector with strong coupling scale $\Lambda$, the axion's \CP~violating interactions have strength 
$ g_s^\chi \sim \theta \Lambda/f_\phi$, 
where $\theta\in [-\pi,\pi]$. (This is the form of the QCD axion's \CP~violating couplings, up to a mild dependence on the masses of the light quarks \cite{Moody:1984ba}.)


\section*{A galactic axion gradient}

As a result of first term in Eq.~\eqref{eq:lag}, DM sources an axion field. We  consider axions with masses $m_\phi \lesssim 1/({\rm pc}) \simeq 10^{-23}~{\rm eV}$ such that this induced axion field extends over galactic scales. 
The equation of motion of $\phi$ in the background of the DM in our galaxy has a time-independent solution
\begin{align}
\label{eq:potential}
    \phi(\vec{r}) = g_s^\chi \int_V d^3{\vec r'} \, \frac{n_{\chi} (\vec{r}') \, e^{- m_\phi | \vec{r}- \vec{r}'| } }{4 \pi |\vec{r} - \vec{r'}|} \, ,
\end{align}
where $\vec{r}$ is relative to the galactic centre and $n_\chi(\vec{r})$ is the DM number density, which we assume to be time-independent and spherically symmetric; corrections to this are relatively small. 
In a full cosmological history one expects $\phi$ to not exactly take the form of the time-independent solution of Eq.~\eqref{eq:potential}. Suppose we start from a spatially homogeneous $\phi$ field in the early Universe (analogous to the pre-inflationary axion scenario), say at $\phi=0$ for convenience. Then, as the galaxy forms, $\phi$'s interaction with dark matter number density will source a non-zero field value. The timescale on which this field is induced can be estimated from the equation of motion and the size of the linear term to be roughly $\left( g_x^\chi n_\chi/f_\phi \right)^{-1/2}$. For all the parameter space that we consider subsequently, this timescale is much shorter than the present-day Hubble time and the galactic formation time, so a non-zero $\phi$ field will always be induced, which we assume reaches the time-independent solution of Eq.~\eqref{eq:potential} before the present-day. We stress however that there could be fluctuations on top of the time-independent solution.\footnote{On cosmological scales these fluctuations would be damped by Hubble expansion but the situation is less clear within the galaxy. Moreover, the distribution of dark matter within the galaxy is changing, which could induce additional fluctuations.}
Because we will consider searches for $\phi$ gradients, any additional such fluctuations would only strengthen the signals that we consider. A detailed study of the full dynamics in the early Universe would be valuable, however it would likely require detailed numerical simulations that go beyond the scope of our present work.

The expression in Eq.~\eqref{eq:potential} simplifies in two limits: First, if the Compton wavelength of the axion, $1/m_\phi$, is shorter than length scale over which the DM density varies, $\lambda_\chi\equiv n_\chi/\left| \boldsymbol{\nabla} n_\chi \right|$, the axion field at any point can be approximated as that sourced from the surrounding $1/m_\phi^3$ volume, which contains $N_{\rm eff}\equiv n_{\chi}/m_\phi^3$ DM particles \cite{Acevedo:2023owd}. As a result, 
\begin{align}
\label{eq:approxpotential}
     \phi(r) \simeq  \frac{g_s^\chi \, N_{\rm eff}  }{\lambda_\phi}  \simeq   \frac{ g_s^\chi\, n_{\chi}(r) }{m_\phi^2}\,.
\end{align}
Second, if the axion's Compton wavelength is comparable to the scale of the galaxy or larger, the axion field is well-approximated by the standard Coulombic potential. 

At the Earth's distance from the galactic centre, $R$, we therefore  have an axion field gradient of
\begin{align} \label{eq:gradphi}
  \left| \boldsymbol{\nabla} \phi(R) \right| \simeq 
\begin{cases}
 g_s^\chi\, \left|\boldsymbol{\nabla} n_{\chi}(R) \right|/m_\phi^2 & \text{if } 1/m_\phi < \lambda_\chi, \\
 g_s^\chi\, N_\chi(R)/R^2 & \text{if } 1/m_\phi > \lambda_\chi\, ,
\end{cases}
\end{align}
where $ N_\chi(R) = 4\pi \int_0^R n_\chi(r) r^2 dr$ is the total number of DM particles within the Earth's galactic orbit.

Such an axion gradient affects the SM particles through the second term in Eq.~\eqref{eq:lag}. In the limit that the SM particles are non-relativistic as is the case in most detectors, the corresponding interaction in the Hamiltonian is (see e.g. \cite{Graham:2017ivz,Alonso:2018dxy} for recent reviews)
\begin{equation}\label{eq:monopole-dipole_Hamiltonian}
    H_{\phi}=-\frac{c_\psi}{f_\phi}\boldsymbol{\nabla} \phi \cdot \mathbf{S}\,,
\end{equation}
where $\mathbf{S}$ is the spin operator of the SM particle.  Consequently, an axion gradient leads to a spin-dependent energy shift $\Delta E$ in SM states, analogous to the Zeeman effect. Using Eq.~\eqref{eq:gradphi}, for a particle with spin aligned with the axion field gradient
\begin{align} \label{eq:DeltaE2}
    \Delta E  & \simeq 
\begin{cases}
  c_\psi g_s^\chi \left| \mathbf{S} \right| \left| \boldsymbol{\nabla} n_{\chi}(R) \right| /(f_\phi m_\phi^2) & \text{if } 1/m_\phi < \lambda_\chi \, , \\
  c_\psi g_s^\chi\left| \mathbf{S} \right| N_\chi(R)/(f_\phi R^2)  & \text{if } 1/m_\phi > \lambda_\chi\,.
\end{cases}   
\end{align}
In other words, there is a monopole-dipole axion force \cite{Moody:1984ba} with the monopole side from the coupling to DM.

Eq.~\eqref{eq:DeltaE2} can easily be evaluated assuming, e.g., an NFW DM profile \cite{Navarro:1995iw}. Fixing the local DM energy density $\rho_0\simeq 0.4~{\rm GeV}/{\rm cm}^3$ and the scale radius to $17.6 \, \rm kpc$, gives
\begin{align}\label{eq:energy_shift}
\Delta E \simeq 3\times 10^{-24}~{\rm eV} \,\kappa 
\begin{cases}
  1/\left(\lambda_\chi m_\phi\right)^2 & \text{if } m_\phi > 1/\lambda_\chi  \, , \\
  1  & \text{if } m_\phi < 1/\lambda_\chi \,,
\end{cases} 
\end{align}
with $1/\lambda_\chi \simeq 4 \times 10^{-28}~ \text{ eV}$ and  
\begin{align} \label{eq:kappa}
  \kappa = \left|\mathbf{S}\right| \left ( \frac{g^\chi_s}{m_\chi/M_{\rm P}} \right )\left ( \frac{10^{9}~\text{ GeV}}{f_\phi/c_\psi}\right )\,,
\end{align}
where $M_{\rm P}$ is the (non-reduced) Planck mass. Not surprisingly the energy shift is largest when $m_\phi \lesssim 1/\lambda_\chi\sim \left(10 \, \rm kpc\right)^{-1}$, such that the induced axion field is sensitive to the DM distribution throughout the entire galaxy.


\section*{Constraints and Experimental Reach}

There are existing, independent, constraints on the interactions of an ultra-light axion with DM and the SM. \CP~violating interactions with DM are bounded because they lead to long-range self-interactions that affect the DM's dynamics \cite{Kesden:2006zb,Desmond:2018euk, Bottaro:2023wkd,Bogorad:2023wzn}. 
Parameterising such forces as 
\begin{align}
    V(r) = -\frac{\alpha \, G_{\rm N} m_\chi^2 }{r} \, \exp{ \left( - r/l \right) } \,,
\end{align}
the constraints are on $\alpha$ as a function of $l$, independent of $m_\chi$. In our theories, $l = 1/m_\phi$ and $g^\chi_s = \sqrt{\alpha} \, m_\chi/M_{\rm P}$, so the combination $g_s^\chi/m_\chi$ is bounded. 
For axion masses of order $10^{-30}~{\rm eV}$, corresponding to super-galactic lengths, these forces must be weaker than gravity, i.e. $g_s^\chi/m_{\chi} \lesssim 1/M_{\rm P}$. For larger axion masses, $m_\phi\gtrsim 1/{\rm kpc}$, the dominant constraints are from observations of the Bullet Cluster and values of $g_s^\chi/m_{\chi}$ that are a few order of magnitude larger than $ 1/M_{\rm P}$ are allowed \cite{Bogorad:2023wzn}. 
Meanwhile, the strongest constraints on \CP~preserving couplings of light axions to SM fermions come from the evolution of stars; the limits on couplings to electrons, $c_e/f_\phi$, and nucleons, $c_N/f_\phi$, are both of order $10^{-9}~\text{GeV}^{-1}$ \cite{Capozzi:2020cbu,Buschmann:2021juv}, see \cite{Caputo:2024oqc} for a recent review. We also note that \CP~violating couplings of the axion to the visible sector (not included in Eq.~\eqref{eq:lag}) must be such that SM monopole-monopole forces are at least a factor of $10^{10}$ weaker than gravity \cite{Berge:2017ovy}.

Remarkably, existing experiments sensitive to spin-dependent energy shifts from a galactic axion gradient explore parts of parameter space that are not excluded by the preceding constraints. 
Unlike measurements of e.g. the muon magnetic dipole moment, the energy shift due to an axion gradient cannot be boosted, so what matters is the \textit{absolute sensitivity}\footnote{By absolute sensitivity we mean that the reach for axion gradients does only depend on the smallest energy shift which can be measured at a given experiment.} of an experiment, which makes extremely precise table-top approaches well-suited. 
Crucially, because the axion gradient couples only to spin, the relative energy shift $\Delta E^{(a)}/\Delta E^{(b)}$ of two different atomic or molecular states ($(a)$, $(b)$) can differ from that due to a magnetic field. This is exploited by so-called co-magnetometers, which use two or more types of fermionic spins to distinguish new-physics energy shifts from uncontrolled magnetic field backgrounds (the precise conditions for general atomic and molecular states are given in \cite{Agrawal:2023lmw}). Co-magnetometry has been used in the past to search for electric dipole moments, axion forces, and more exotic signals, see \cite{Safronova:2017xyt,Terrano:2021zyh} for reviews.

The radial orientation of the axion gradient relative to the galactic centre offers additional experimental opportunities. For a fixed experiment, it makes the induced energy shift a pseudo-DC signal with daily modulation due to the Earth's rotation. This feature of the signal is different to an axion gradient generated by the Earth \cite{Moody:1984ba,PhysRevLett.68.135,PhysRevD.78.092006}, see  \cite{Fedderke:2023dwj,Agrawal:2022wjm,Agrawal:2023lmw,Davoudiasl:2022gdg} for recent discussions and \cite{Crescini:2017uxs,Crescini:2020ykp,Arvanitaki:2014dfa} for related studies of forces from test masses.  
Moreover, by rotating an experiment additional modulation can be induced. In case of a signal, the sharp prediction of the orientation of the axion gradient would allow discrimination from possible neglected backgrounds or other new physics sources.

Out of the current experiments, those of Refs.~\cite{PhysRevLett.105.151604} and \cite{Smiciklas:2011xq} are among the most sensitive to a galactic axion gradient. The proposal in \cite{Wei:2023rzs} may also be easily adapted to perform these measurements. These use co-magnetometers based on K-$^3$He and $^{21}$Ne-Rb-K to reach frequency resolutions at the level of $0.7~{\rm nHz}$ and $0.5~{\rm nHz}$ respectively, corresponding to energies of order $10^{-24}~{\rm eV}$. Such experiments  have been used to search for Lorentz violation of extra-solar origin (e.g. as in the Kostelecky SM extension \cite{PhysRevD.58.116002}), making use of a rotating platform to help remove backgrounds and systematics. 
The resulting limits can immediately be reinterpreted as constraints on our theories. Moreover,  experiments based on $^{21}$Ne-Rb-K \cite{Smiciklas:2011xq} (see also \cite{Wei:2023rzs}) are particularly promising for future improvements, having obtained a sensitivity similar to the K-$^3$He experiment with a factor of $8$ less integration time. Indeed, in Ref.~\cite{PhysRevLett.105.151604}, it is mentioned that with future upgrades, resolution at the level of $10^{-3}~{\rm nHz}$ might be achieved. There are also several other experiments and proposals with sensitivities to energy shifts at the level of $\mathcal{O}({\rm nHz})$ \cite{Alexander:2022rmq,Suleiman:2021whz,Brandenstein:2022eif}.
The resulting limits can immediately be reinterpreted as constraints on our theories. Moreover,  experiments based on $^{21}$Ne-Rb-K \cite{Smiciklas:2011xq} (see also \cite{Wei:2023rzs}) are particularly promising for future improvements, having obtained a sensitivity similar to the K-$^3$He experiment with a factor of $8$ less integration time. Indeed, in Ref.~\cite{PhysRevLett.105.151604}, it is mentioned that with future upgrades, resolution at the level of $10^{-3}~{\rm nHz}$ might be achieved. There are also several other experiments and proposals with sensitivities to energy shifts at the level of $\mathcal{O}({\rm nHz})$ \cite{Alexander:2022rmq,Suleiman:2021whz,Brandenstein:2022eif}.


\section*{Underlying theories}

So far we have taken Eq.~\eqref{eq:lag} at face value, however this can miss important effects that arise in a complete theory. In particular, such a Lagrangian is expected to appear as the first terms in an expansion in an axion field $\phi \ll f_\phi$, and higher order terms will be relevant if values $\phi\simeq f_\phi$ are induced in the galaxy. At this point, given that axions have a  compact field range, the interaction $g_s^\chi \phi \bar{\chi} \chi$ should be replaced by $h(\phi/f_\phi)\bar{\chi} \chi$ where $h$ is a function with period $2\pi$ (similarly, the axion mass will be replaced by a periodic potential). We do not consider higher-order derivative interactions since they are generally shift-symmetric and will not source an axion gradient.

It can immediately be seen that the compactness of the axion field is going to be borderline relevant for values of $g_s^\chi$ that lead to detectable energy shifts. Regardless of the axion mass, the maximum galactic axion field gradient possible without the compactness of the axion field being an issue is $|\boldsymbol{\nabla} \phi| \lesssim f_\phi/R$ such that 
\begin{equation}\label{eq:field_range}
   \Delta E \lesssim \frac{c_\psi}{f_\phi} \frac{f_\phi}{R} \simeq c_\psi 10^{-27}~{\rm eV}  \,.
\end{equation}
For $c_\psi\sim\mathcal{O}(1)$ this is a couple of orders of magnitude below current sensitivities and might be achievable in the future. Assuming $m_\phi \lesssim 1/\lambda_\chi $ for simplicity, using Eq.~\eqref{eq:potential}, in the Milky Way $\phi\sim f_\phi$ is reached for 
\begin{equation} \label{eq:gsvals}
g_s^\chi= 10^{-26} \left(\frac{m_\chi}{{\rm MeV}}\right)\left( \frac{f_\phi}{10^{9}~{\rm GeV}}\right)~,
\end{equation}
which is consistent with DM-DM self-interaction constraints, that is $g_s^\chi / m_\chi < 1/ M_P$, independently of the DM mass for $f_\phi\geq 10^{5}~{\rm GeV}$. With the parameterization $g_s^\chi=\theta\Lambda/f_\phi$, Eq.~\eqref{eq:gsvals} corresponds to 
\begin{equation} \label{eq:theta}
\theta = 10^{-4} \left(\frac{10^{-4}~{\rm eV}}{\Lambda}\right)\left(\frac{m_\chi}{{\rm MeV}} \right)\left( \frac{f_\phi}{10^9~{\rm GeV}} \right)^2~,
\end{equation}
i.e. only a small CP violating angle in the hidden sector if $m_\chi$ is not too large. We explain the choice of normalisation of $\Lambda$ shortly.

In the case of larger $g_s^\chi$,  so additional operators are relevant, the induced axion field typically saturates at values of order $f_\phi$ rather than continuing to grow. Fields $\phi \gg f_\phi$ are not forbidden by the axion field range's compact nature; this would simply correspond to the axion winding its fundamental domain. Instead, saturation occurs because $\phi$ reaches a value such that $\partial_\phi h(\phi/f_\phi)=0$ at which point the coupling that sources a $\phi$ field turns off, effectively \textit{self-screening} the DM. For instance, a typical form of the full axion potential might be (further details are given in Appendix~\ref{app:compact})
\begin{equation} \label{eq:CPVfull}
V\supset -\Lambda^4 \cos\left( \phi/f_\phi \right) + \bar{\chi} \chi \Lambda \cos(\phi/f_\phi+\theta)~.
\end{equation}
Expanded at small $\phi/f_\phi$ the second term leads to a \CP~violating coupling $\sin(\theta) \frac{\Lambda}{f_\phi} \phi  \, \bar{\chi}\chi$. As $\phi/ f_\phi \simeq \pi-\theta $ is approached in the galaxy, a non-zero $\bar{\chi}\chi$ background contributes less and less to the equation of motion of $\phi$. Eventually, the induced galactic axion field saturates once $\phi/f_\phi\simeq \pi-\theta$ is reached. We also note that, if $\theta\ll 1$, 
as the background $\phi/ f_\phi$ increases 
the effective coupling of $\phi$ to $\chi$ is first enhanced, reaching a maximum strength at $\phi/ f_\phi \simeq \pi/2-\theta$, before being screened.

We therefore identify $\Delta E\lesssim 10^{-27}~{\rm eV}$, arising from minimal theories, as a well-motivated target for future experiments. However, we also note there are various ways that larger $\Delta E$ can arise from consistent UV theories. Most simply, values $c_\psi\sim 100$ are possible in normal models. Meanwhile, more unusual UV completions such as clockwork theories \cite{Kaplan:2015fuy,Farina:2016tgd,Agrawal:2018mkd} can lead to effective values of $c_\psi$ that are exponentially large. Finally, if the axion potential has a monodromy   \cite{Silverstein:2008sg,McAllister:2008hb,Kaloper:2008fb} the induced axion field might not saturate at values of order $f_\phi$.

Another question that depends on the particular theory underlying Eq.~\eqref{eq:lag} is whether an extremely small axion mass, $m_\phi\sim 10^{-27}~{\rm eV}$, is possible without fine tuning. 
If $g_s^\chi$ indeed arises from a hidden sector gauge group running into strong coupling at scale $\Lambda$, a typical expectation is $m_\phi\sim \Lambda^2/f_\phi$ (a smaller $m_\phi$ is possible if there are light hidden sector chiral fermions). To obtain $m_\phi\sim 10^{-26}~{\rm eV}$ for $f_\phi\sim 10^{9}~{\rm GeV}$ then requires $\Lambda\lesssim 10^{-4}~{\rm eV}$. 

Small values of $\Lambda$ are not necessarily problematic, although in the early universe the hidden sector must be colder than the SM to avoid bounds on new relativistic degrees of freedom \cite{Aghanim:2018eyx}. However, observations require that fermionic DM must have a mass much larger than such $\Lambda$ \cite{Tremaine:1979we,Boyarsky:2008ju}. This is also needed so that the galactic DM number density is smaller than $\Lambda^{3}$ and the hidden sector is not locally deconfined.  Moreover, $m_\chi > \Lambda$ guarantees that the DM mass induced by a background $\phi \lesssim f_\phi$ is smaller than the bare DM mass, both for $g_s=\theta\Lambda/f_\phi$ (with $\theta\lesssim 1$) and the example completion in Eq.~\eqref{eq:CPVfull}.

Whether a DM candidate with a mass larger than $\Lambda$ gets substantial \CP~violating interactions with the axion is model-dependent. 
In Appendix~\ref{app:DM}, we describe an example theory in which DM is a particle in the adjoint of a hidden sector gauge group that forms bound states with dark gluons, as studied in Ref.~\cite{Falkowski:2009yz,Boddy:2014qxa,Boddy:2014yra,Contino:2018crt}. Provided the mass of the adjoint field is greater than roughly a ${\rm keV}$ such a state can be a viable DM candidate, with the required \CP~violating interactions arising from the gluon part of the bound state.


\section*{Results}

\begin{figure}[ht!]
\includegraphics[width=0.995\columnwidth]{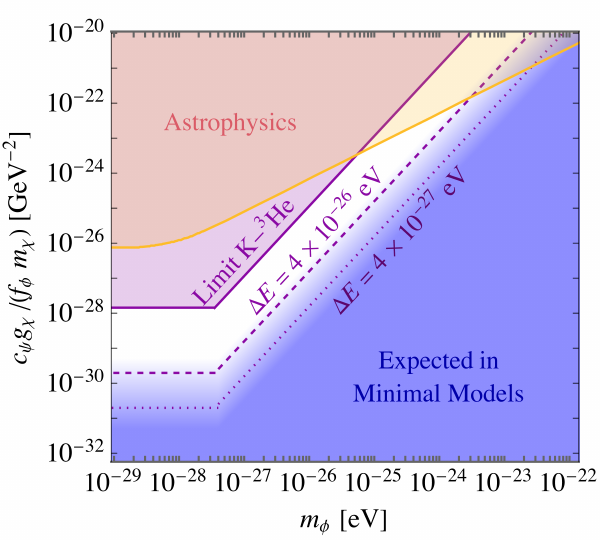}
\caption{Existing constraints  from DM self-interactions (labelled ``Astrophysics'') and our new limit (labelled ``Limit K-$^3$He''; obtained by reinterpreting the results of Ref.~\cite{PhysRevLett.105.151604}) on theories that lead to a galactic axion gradient. Results are shown as a function of the parameters in the effective Lagrangian in Eq.~\eqref{eq:lag}. We also  plot the reach of plausible future experiments with sensitivity to spin-dependent energy shifts of $\Delta E=4\times 10^{-26}~{\rm eV}$ and $4\times 10^{-27}~{\rm eV}$. Finally, we indicate the region of parameter space in which Eq.~\eqref{eq:lag} is self-consistent in minimal UV completions (labelled ``Expected in Minimal Models'').}
\label{fig:parameterspace}
\end{figure}

In Figure~\ref{fig:parameterspace}, we show the parameter space of theories with a galactic axion gradient in the plane of $(c_\psi g_s^{\chi})/(f_\phi m_\chi)$ against $m_\phi$. In this, we plot the existing constraints from astrophysics by saturating the allowed values of $c_\phi/f_\phi$ and $g_s^\chi/m_\chi$  separately; any point in parameter space outside this region (labeled as "Astrophysics" in Fig.\ref{fig:parameterspace}) corresponds to a theory that is consistent with existing constraints.  Meanwhile, as can be seen from Eqs.\eqref{eq:energy_shift} and \eqref{eq:kappa}, the induced spin-dependent energy shift of SM fields depends on the couplings via the combination plotted. We show our new limit, obtained by reinterpreting the results of Ref.~\cite{PhysRevLett.105.151604}, and also the reach of possible future experiments with plausible improvements in sensitivity to $\Delta E$. These curves are obtained from the analytic results in Eq.~\eqref{eq:energy_shift}; in reality the transition around $m_\phi\simeq 1/\lambda_\chi$ will be smooth. 

We also indicate in the same Figure the part of parameter space for which the galactic axion field induced by Eq.~\eqref{eq:lag} is self-consistently such that the compactness of the axion field can be neglected. In particular, we fix $g_s^\chi/m_\chi$ such that the $\phi = f_\phi$ at the galactic centre (assuming the underlying theory is such that a $\phi$ background does not self-enhance the $\phi$--$\chi$ coupling, as is e.g. the case for $\theta\simeq 1$ in Eq.~\eqref{eq:CPVfull}). Given that the induced $\phi$ is proportional to $g_s^\chi/m_\chi$, the maximum allowed $g_s^\chi/m_\chi$ is proportional to $f_\phi$, such that the combination $g_s^\chi c_\psi/(f_\phi m_\chi)$ is independent of $f_\phi$, but depends on $m_\phi$ and $c_\psi$. The value of $c_\psi$ is model dependent and the maximum typical value not sharply determined, so we blur the edge of this region over the range corresponding to $1<c_\psi<100$. 
We stress that values of $g_s^\chi$ that do not saturate $\phi\simeq f_\phi$ (and smaller $c_\psi$) are equally plausible and lead to theories within the blue ``Expected in Minimal Models'' region. 

For $m_\phi\lesssim 10^{-24}~{\rm eV}$, existing and future experiments can probe sizable parts of parameter space of the effective Lagrangian in Eq.~\eqref{eq:lag} that are not otherwise excluded. As expected, for fixed $(c_\psi g_s^\chi)/(f_\phi m_\chi)$ the signal is weaker at larger axion masses, but the observational constraints also weaken. Although beyond current experimental sensitivity, future experiments will begin to explore the parameter space for which the galactic axion field is smaller than $f_\phi$ and the effective Lagrangian Eq.~\eqref{eq:lag} is valid even for the simplest underlying theories. We also reiterate that more exotic theories (with effective $c_\psi \gg 1$) can lie in the part of parameter space above this that is already experimentally tested. 
Finally, we note that for all $m_\phi$ and  $f_\phi$ considered, the DM relic abundance of $\phi$ itself is indeed negligible assuming misalignment production.

\vspace{-0.3cm}


\section*{Discussion}

The theories we have studied have an elegant UV interpretation: It is plausible that a typical string theory compactification leads to multiple light axions and several dark sectors. Some of the axions might have \CP~violating couplings to the dark sectors. One of these dark sectors could host the DM and would communicate with the visible sector through a long-range axion force in the form of a galactic axion gradient.  
Ultra-precise table top experiments already have impressive sensitivity to this scenario and orders-of-magnitude improvement is possible in the future.  
Such experimental searches are ``looking under a lamppost'', in that a theory need not be in the upper part of the ``minimal theories'' parameter space shaded in Figure~\ref{fig:parameterspace} that can be explored in the near-future. Nevertheless, given that such technologies are being developed for independent purposes it is valuable to know that experimental results can be reinterpreted as a search for DM. Moreover, the prediction that the energy shift is maximised when the SM particle's spin is aligned with the galactic centre (typically not expected for other new-physics signals) might allow for simple experimental adaptions to increase sensitivity, e.g. by reorientating existing apparatus.

Although the DM \CP~violating/ SM \CP~preserving axion portal that we have considered is novel, we note that related ideas have been analysed in the past. Experiments searching for monopole-dipole forces where both parts of the interaction correspond to SM fermions have been performed \cite{Lee:2018vaq,Crescini:2020ykp} and there are promising prospects for the near future \cite{Arvanitaki:2014dfa,Alexander:2022rmq,Agrawal:2022wjm,Agrawal:2023lmw,Suleiman:2021whz,Brandenstein:2022eif,Fan:2023hci}.

We have given an example, simple, UV completion in which the DM is a bound state of a dark fermion and a dark gluon. However, there are likely to be other models that also reduce to Eq.~\eqref{eq:lag} at low energies, and it would be interesting to explore whether these can lead to new phenomenology or signals. For example, there might be suitable theories in which the DM is a weakly coupled bound state of the type $\chi=\mathcal{QQ}$ (where $\mathcal{Q}$ is a heavy hidden sector quark) similar to bottomonium in QCD. There might also be UV completions in which the induced axion field is not screened if $\phi\sim f_\phi$ is reached in the galaxy.  
 A more detailed study of the cosmology of these scenarios, including mechanisms to get the correct relic abundance as well as studying their effect in other cosmological and astrophysical observables, lies beyond the scope of this work, but would be an valuable direction for future study.

We have focused on the time-independent solution for the sourced $\phi$ field, Eq.~\eqref{eq:potential}, which we expect to be, approximately, reached in the Milky Way after a full cosmological history. However, it would be interesting if $\phi$ was not fully relaxed to this form yet, e.g. due to oscillations left over from galactic formation, or had time-dependence, e.g. due to the dynamics of DM substructure within the galaxy. As mentioned, such effects could lead to stronger signals, and it would be interesting to investigate them in the future. 
More generally, we have conservatively applied constraints on long range DM-DM forces ignoring the periodicity of the axion potential, and it would be interesting to reanalyse these. Additionally, as mentioned in Ref.~\cite{Bogorad:2023wzn}, it would be interesting to investigate the cosmological effects of ${\rm kpc}$ range DM-DM self-interactions studied in this paper, similarly to the analysis carried out in Refs.\cite{Gradwohl:1992ue,Nusser:2004qu,Hellwing:2008qf,Bottaro:2023wkd} for forces with longer range.  These might lead to stronger constraints than the existing constraints we have made use of as well as possible complementary signals. This interesting direction is left for future work.

\begin{acknowledgments}
We thank Prateek Agrawal for collaboration at early stages of this project as well as for very useful discussions. 
We also thank Junwu Huang and Surjeet Rajendran for useful discussions, and Javier Fernandez Acevedo and Junwu Huang for useful comments on a draft. 
EH acknowledges the UK Science and Technology Facilities Council for support through the Quantum Sensors for the Hidden Sector collaboration under the grant ST/T006145/1 and UK Research and Innovation Future Leader Fellowship MR/V024566/1.
MRs work is supported by the STFC grant ST/T006242/1. MR thanks the CERN Theory Department, where this work was finished, for hospitality.
\end{acknowledgments}

\appendix

\section{Compactness of the axion field}\label{app:compact}

As described in the main text, the fact that the axion couplings are periodic means that the DM is automatically screened and the induced galactic axion field saturates once values $\phi\sim f_\phi$ are reached (rather than winding the fundamental domain and growing indefinitely). To see this explicitly, we consider an axion potential motivated by that of the QCD axion, including its CP violating couplings to quarks. 

Suppose that the axion potential is generated by a hidden sector gauge group running into strong coupling in the IR. In the UV, we assume the axion is coupled to the dark gluons as:
\begin{equation}\label{eq:dark_confining_sector}
    \frac{\alpha_D}{f_\phi}\phi G_D\tilde{G}_D\,.
\end{equation}
Under quite general conditions, the minimum of the contribution to the potential resulting from strong coupling is CP conserving \cite{Vafa:1983tf}. In the absence of light chiral fermions in the hidden sector (which could lead to a chiral suppression of the axion mass, analogous to the QCD axion mass' dependence on the light quark masses), the resulting axion mass is of order 
$m_\phi\sim 
\Lambda^2/f_\phi$.

Let us moreover assume that there is an additional \CP~violating contribution to the axion's potential, associated with a scale $\Lambda'$, that shifts the minimum slightly. Temporarily changing conventions such that $\phi=0$ is the CP conserving point, the resulting axion potential  (considering a simple analytic functional form; more complicated potentials lead to the same effects) is \cite{Moody:1984ba,Bertolini:2020hjc}
\begin{equation}
V= -\Lambda^4 \cos\left(\frac{\phi}{f_\phi}\right) -\Lambda'^4 \cos\left(\frac{\phi}{f_\phi} + \delta \right) ~.
\end{equation}

Analogous to the QCD axion, if the DM is charged under the dark gauge group we expect a CP violating coupling of the axion to dark matter of the form
\begin{equation} \label{eq:cpva}
V\supset  \Lambda \bar{\chi}\chi \cos\left(\frac{\phi}{f_\phi}\right)~.
\end{equation}
In more detail, a coupling of simply $\Lambda$ in Eq.~\eqref{eq:cpva} is expected for DM with a mass of order $\Lambda$ and for the quark-gluon bound state described in the main text. For a $\mathcal{QQ}$ bound state with mass larger than $\Lambda$ the coupling is expected to take a more complicated form. 
If there is a chiral suppression of the axion mass, the strength of the interaction in Eq.~\eqref{eq:cpva} is typically also suppressed (an analogous dependence on $m_um_d$ appears in the QCD axion's \CP~violating interactions).

Redefining $\phi \mapsto \phi - f_\phi \theta$ with $\theta$ such that (with  $\left<\bar{\chi}\chi\right>=0$) the minimum of the potential is at $\phi=0$, the matter part of the potential takes the form
\begin{equation} \label{eq:CPVfullb}
V\supset \Lambda\bar{\chi} \chi  \cos(\phi/f_\phi+\theta)~.
\end{equation}
Expanded at small $\phi/f_\phi$  this leads to the required CP violating coupling:
\begin{equation}
    \mathcal{L}\supset \sin(\theta) \Lambda/f_\phi \phi  \, \bar{\chi}\chi\,.
\end{equation}

 From Eq.~\eqref{eq:CPVfullb} it can be seen that as $\phi/ f_\phi \simeq \pi-\theta$ is approached, a non-zero $\bar{\chi}\chi$ background contributes less and less to the equation of motion of $\phi$ due to the non-linearities. Numerical, time-independent, solutions of $\phi$'s equation of motion in the background of a static galactic DM distribution (with boundary condition $\phi=0$ at spatial infinity) indeed shown that for $\Lambda/f_\phi$ sufficiently large that $\phi\sim f_\phi$ is reached, the axion field indeed saturates.

This is consistent with the results in \cite{Hook:2017psm} (see also \cite{Balkin:2022qer,Balkin:2023xtr}), where it is shown that for a light QCD axion -- that is, an axion coupled to QCD with a potential that is suppressed with respect to the standard prediction --  the finite density effects inside neutron stars can make the axion field to sit near $a/f_a=\pi$. In this case, the shift in axion field value is only due to the finite density potential, which has a relative $\pi$ phase with respect to the vacuum potential, and is much more relevant than the CPV contributions for the QCD axion.

\section{The axion mass and a model of dark matter}\label{app:DM}

Give the extremely light axions that we have considered, it is interesting to see that this can be achieved without fine-tuning, although of course one can simply      
disregard fine-tuning arguments, which opens up a wide range of possible DM models.  
Assuming, as before, that the axion potential is dominantly generated by a hidden sector running into strong coupling, this requires $\Lambda \lesssim 10^{-3}~{\rm eV}$ assuming $f_\phi\sim 10^9~{\rm GeV}$. It also requires that there are no large contributions to the axion's potential from the high energy scale; this is a UV dependent issue analogous to the QCD axion quality problem, which we do not consider further.

Insisting on such small values of $\Lambda$, restricts the range of phenomenologically viable DM candidates that can also have sizable \CP~violating interactions with the axion. In particular, for fermionic DM the Tremaine-Gunn bound \cite{PhysRevLett.42.407} requires $m_\chi\gtrsim {\rm keV}\gg\Lambda$. Additionally, to avoid the hidden sector gauge group being in the deconfined phase in the galaxy requires the typical inter-DM spacing $n_{\chi}^{-1/3} \gg \Lambda^{-1}$. Given that the DM energy density in the galaxy is of order $10^{-6}~{\rm eV}^4 \gg \Lambda^4$ this again requires a DM mass much larger than $\Lambda$.

DM candidates with a mass much larger $\Lambda$ might naively be expected to have \CP~violating axion couplings that are strongly suppressed, and therefore do not lead to a large enough galactic axion gradient to obtain a detectable signal, but this is not necessarily the case. As an example, suppose that the hidden sector contains a dark fermion with a mass much larger than the dark confinement scale, $m_\mathcal{Q}\gg \Lambda$, in the adjoint representation, $\mathcal{Q}\sim \text{Adj}$. Then the DM $\chi$ can consist of bound states of the fermion, $\mathcal{Q}$, and a dark gluon, $g$. This kind of  dark matter candidate has been proposed in Ref.~\cite{Falkowski:2009yz,Boddy:2014qxa,Boddy:2014yra,Contino:2018crt}, see also \cite{DeLuca:2018mzn} for a related discussion of a similar DM candidate charged under the SM $SU(3)$.

The mass of such DM bound states, $\chi\sim  \mathcal{Q}g$, is dominated by the constituent quark
$   m_{\chi}= m_\mathcal{Q}$, while its size is set by the confinement scale, $r\sim \Lambda^{-1}\gg m_\mathcal{Q}^{-1}$. This implies that despite $\rho_{DM}\gg \Lambda^4$, the typical occupancy number is low and, for sufficiently large $m_\chi$, we will be in the confined phase of the theory provided $m_\chi \gg {\rm keV}$ for $\Lambda=10^{-3}~{\rm eV}$. 
\CP~violating interactions between the DM and the axion are dominantly induced by the mixing between the axion and the lightest \CP-even glueball. First, note that the the mixing angle between the axion and the lightest \CP-odd glueball can be estimated in analogy with the axion-pion mixing and is of order $\Lambda/f_\phi$. Then, in the presence of a non-zero $\theta$, the CP-odd and CP-even glueballs will mix, which leads to an effective CPV coupling of the axion to the DM bound state of order $ g_s^\chi\sim \theta \Lambda/f_\phi$, which is the parametric form assumed in the main text.

Finally, we note that even assuming thermal production, the calculation of the relic abundance in this scenario is complicated, involving many different regimes and effects as well as non-perturbative dynamics  \cite{Contino:2018crt}. Moreover, it is possible that the relic abundance is set by an initial DM asymmetry. Consequently, for our present work, we do not attempt to analyse the full cosmological history or specify a complete theory that gives the correct DM relic abundance consistent with all observational constraints. We do however note that for small $\Lambda$ it is likely to be required that the dark sector is colder than the visible sector in the early universe. This is needed both to satisfy observational constraints on additional relativistic degrees of freedom, and such that the dark sector is in a confined phase when the cosmic microwave background forms so that the dark matter does not have strong, long-range, self-interactions at these times.


%

\end{document}